\newif\ifabstract
\abstracttrue
 \abstractfalse 
\newif\iffull
\ifabstract \fullfalse \else \fulltrue \fi

\documentclass[11pt]{article}
\usepackage{amsfonts}
\usepackage{amssymb}
\usepackage{amstext}
\usepackage{amsmath}
\usepackage{xspace}
\usepackage{theorem}
\usepackage{graphicx}
\usepackage{url}
\usepackage{graphics}
\usepackage{colordvi}
\usepackage{colordvi}
\usepackage{subfigure}

\usepackage{natbib}
\usepackage{wasysym}

\usepackage{moreverb} 
\usepackage{textcomp} \usepackage{epsfig}
\usepackage{color} \usepackage{upgreek}
\usepackage{bbm}

\usepackage{subeqnarray}

\graphicspath{{Figures/}}
\pdfoutput=1
\usepackage{natbib}

\advance \oddsidemargin by -0.8in
\newcommand{\myparskip}{3pt}
\parskip \myparskip

\newsavebox{\astrutbox}
\sbox{\astrutbox}{\rule[-5pt]{0pt}{20pt}}

\begin{document}

\title{Experimental investigation of nonlinear internal waves in deep water with miscible fluids}
\author{Roberto Camassa \thanks{Carolina Center for Interdisciplinary Applied Mathematics, Department of Mathematics, University of North Carolina, Chapel Hill, NC 27599, USA. Email: \tt{camassa@amath.unc.edu}}
\and Matthew W. Hurley\thanks{Carolina Center for Interdisciplinary Applied Mathematics, Department of Mathematics, University of North Carolina, Chapel Hill, NC 27599, USA. Email: {\tt hurleymatthew1@gmail.com}.}
\and Richard M. McLaughlin\thanks{Carolina Center for Interdisciplinary Applied Mathematics, Department of Mathematics, University of North Carolina, Chapel Hill, NC 27599, USA. Email: {\tt rmm@email.unc.edu}.}
\and Pierre-Yves Passaggia\thanks{Department of Marine Sciences, University of North Carolina, Chapel Hill, NC 27599, USA. Email: {\tt passaggia@unc.edu}.}
\and Colin F. C. Thomson\thanks{Carolina Center for Interdisciplinary Applied Mathematics, Department of Mathematics, University of North Carolina, Chapel Hill, NC 27599, USA. Email: {\tt cfct@email.unc.edu}.}
}

\maketitle

\thispagestyle{empty}

\begin{abstract}
Laboratory experimental results are presented for nonlinear Internal Solitary Waves (ISW) propagation in `{\it deep water}' configuration with {\it miscible} fluids. The results are validated against direct numerical simulations and traveling wave exact solutions where the effect of the diffused interface is taken into account. The waves are generated by means of a dam break and their evolution is recorded with Laser Induced Fluorescence (LIF) and Particle Image Velocimetry (PIV).
In particular, data collected in a frame moving with the waves are presented here for the first time.
Our results are representative of geophysical applications in the deep ocean where weakly nonlinear theories fail to capture the characteristics of large amplitude ISWs from field observations.
\end{abstract}


\section{Introduction}\label{sec:intro}
Diurnal and semi-diurnal ocean tides generate large amplitude ISWs, as observed for instance on the Hawaiian ridge \citep{Rudnick:03}, near the Luzon strait \citep{Duda:04,Alford:15}, in the gulf of Alaska \citep{churnside:05}, in the Sulu sea \citep{Apel:85}, off the coast of California \citep{Pinkel:79}, and more recently in the Tasman sea \citep{Johnston:15}. These large amplitude ISWs propagate for very long distances from their sources \citep{Hosegood:06,Kunze:12,simmons2012simulating} across the deep ocean (see \citet{HelfrichM:06} for a review on ISWs), have been observed to reach amplitudes as large as $300$m \citep{Rudnick:03} with nonlinearity $\alpha = \eta_{\rm{max}}/h \approx 5$, where $h$ is an appropriate reference depth scale \citep{Stanton:98}. Such large amplitude internal solitary waves induce localized regions of strong shear flows that contribute to a large amount of mixing near the bottom \citep{Kunze:12,vanHaren:13} and across the thermocline \citep{Liu:85,brandt:02}. Internal solitary waves are also likely to be amplified by interaction with  mesoscale eddies \citep{xie2015simulations}. 
ISWs play a central role in ocean circulation as they provide a means for mixing and dissipation in the pycnocline of both the deep ocean \citep{Grimshaw:03,grimshaw2012effect} and the continental shelf \citep{Lamb:14,Alford:15,bourgault:16,passaggia2018optimal}.

While the dynamics of ISWs in shallow configurations has arguably  received most of the attention in the literature (see, e.g., \citet{HelfrichM:06} for a review, and \citet{Ertekin16} very recent developments) its deep water counterpart has been extensively studied mainly from a mathematical perspective, and this mostly in the weakly nonlinear and two-layer regimes modelled by, respectvely,  the Korteweg-de-Vries (KdV) \citep{Grimshaw:81}, the Intermediate Long Wave (ILW) \citep{DavisA:67,Kubota:78} and the Benjamin-Ono (B-O) equations \citep{Benjamin:1986,KalischB:00}. More recently, \citet{Choi:1999} derived a fully-nonlinear model for internal waves in two-layer deep water configuration that makes no assumptions on the amplitude of the wave. The system of equations in the depth and the depth-averaged fluid velocity of the thin layer reduces the ILW and BO models in the appropriate weakly nonlinear, unidirectional limit, but can accommodate waves in the nonlinear regime as well. 
Mathematically, the intermediate long wave equation and its infinite-depth limit, the Benjamin-Ono equation, are completely integrable models for internal waves in deep water, and in principle allow for exact solutions to be computed. While these are certainly a desirable property of a physical model, the weak nonlinearity assumption excludes internal wave phenomena commonly seen in nature. In particular, wave amplitudes are observed to be of magnitudes several times the thickness of the upper layer of water, which falls well outside the weak nonlinearity assumption.
The next step towards an accurate comparison can consist of modeling ISWs by solving the Dubreil-Jacotin-Long (DJL) equation \citep{dubreil1934determination,long1953some}, which gives steady-state solutions to the Euler equations in the moving frame of the wave. While this approach does not have the appeal of analytical-type methods previously cited, it is not restricted to simple stratifications and provides an intermediate theoretical tool between direct, time-
dependent numerical simulations and integrable or analytically tractable models.

Experimental and field observations are routinely compared with numerical solutions to support results extracted from field experiments and extrapolate integral quantities such as wavelength and wave speed \citep{vlasenko2000structure,Preusse:12,Preusse2012seasonal,Lien:14}. Recently, \citet{passaggia2018optimal} used a DJL solution to compute waves corresponding to the field measurements of \citet{Moum:03} and provided an accurate representation of both waves and instability dynamics. Among the open problems in ISW modelling, the role played in the dynamics by turbulent wakes not fully separated by the laminar wave motion   is yet to be studied in detail. In what follows, we indirectly address this issue by a systematic comparison of experimental data with analysis of the DJL equations and direct numerical simulations, with solutions and observations of waves both close and far from their generation site.

Experiments in the miscible and shallow regime have been extensively carried out (see, e.g., \citet{Grue:97,Grue:1999,Carr:08,Fructus:09,Carr:11,Carr:17}).  However, for the deep water regime, results of interfacial waves from experiments mostly used immiscible fluids (e.g. using water/petrol or water/silicon oil) see, e.g.,  \citet{Michallet:98,Kodaira:16}.
To the best of our knowledge, the results presented herein are the first to be performed with miscible fluids on much larger scales than other experiments, both for the horizontal and/or the vertical length scales. The use of salt as a stratifying agent, plus the size of the domain, positively indicates that the results are representative across several length scales and indicate scalability to the extreme case of oceanic ISWs. This assertion is confirmed in the conclusions of the present study,  where our experimental, numerical and theoretical results are compared with observations from field measurements taken near the Luzon strait for large amplitude internal solitary waves \citep{Ramp04,Huang16}.

In order to compare wave properties, we collect local profiles of fluid velocities and density from Particle Image Velocimetry (PIV) and Laser Induced Fluorescence (LIF); these tools allow computation of amplitude, speed, and wavelengths from a dataset. These properties are then compared with DJL solutions and direct numerical simulations of the Euler equations. We show that experiments and simulations do capture the dynamics of internal solitary waves, and, in particular, we show that experiments closely reproduce the DJL predictions, thus also assessing the effect of a diffused interface. Direct numerical simulations of two-dimensional time evolution governed by the Euler equations are in good agreement with wave quantities. In contrast with DJL solutions, the time dependent wave profiles clearly highlight a lack of fore-aft symmetry,  which could be caused by dynamical interaction of the emerging (main) solitary wave with the wake from the generation site for traveling distances commensurate to those of the experiments. 

Specifically, the paper is organized as follows:
%
Section \S~\ref{sec:exp_setup} presents a description of the experimental facility and details about the methods used to perform wave measurements. 
The numerical strategies are reported in \S~\ref{sec:num_methods} together with a description of the initial conditions used to replicate the dam-break problem. The results are summarized in \S~\ref{sec:results} and discussed in \S~\ref{sec:discussion} with respect to theoretical models predictions and discussed in perspective to field field  measurements.

\section{Experimental setup}\label{sec:exp_setup}


\begin{figure*}[t!]
\centering
(a)\scalebox{0.4}{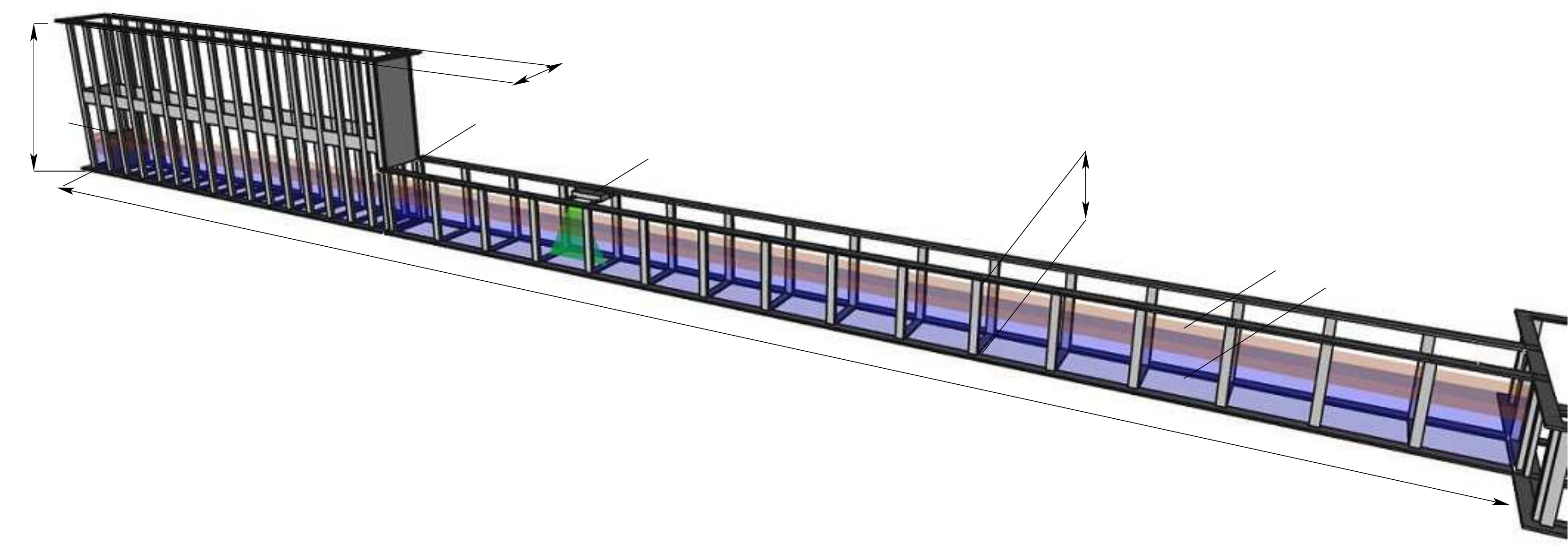}
(b)\hspace{-10mm}\scalebox{0.38}{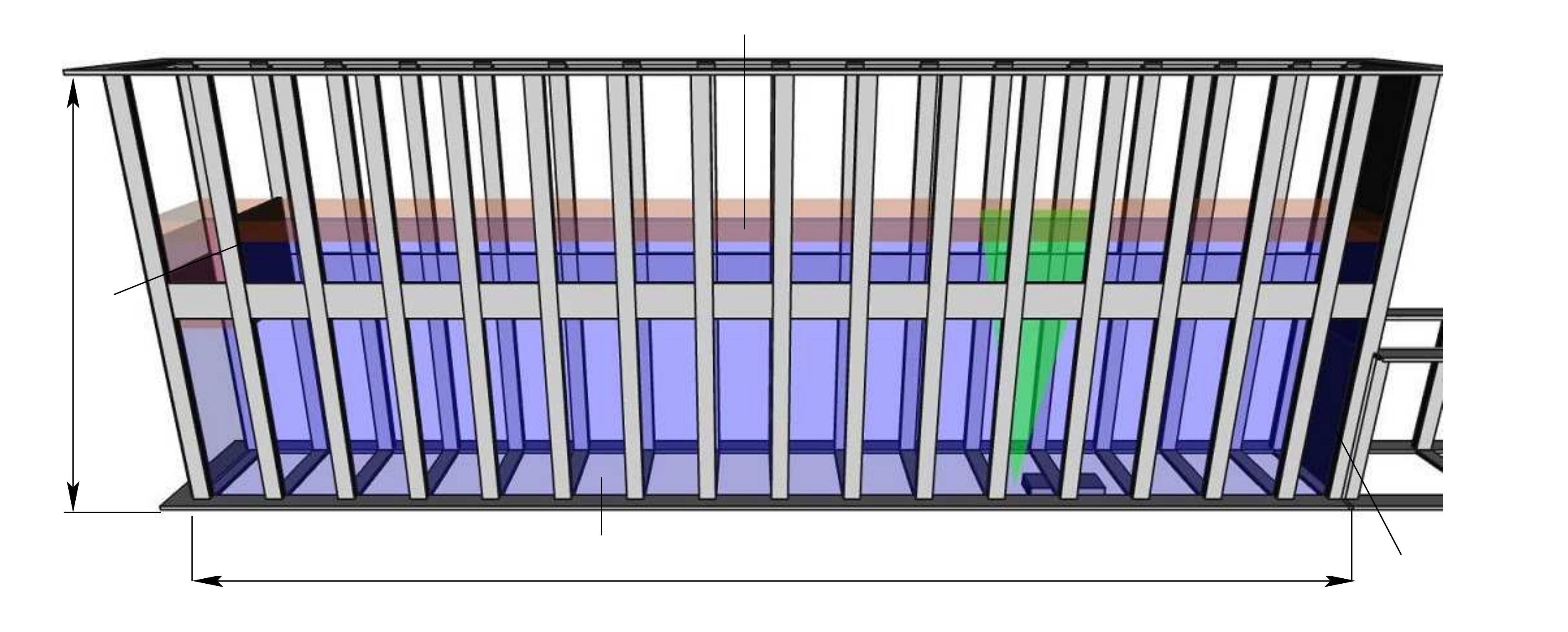}
\caption{Schematic showing the two section of the of the modular wave tank for the long setup (a) and the deep setup (b).}
\label{CAD} 
\end{figure*}

Experiments were conducted at the Joint Applied Mathematics and Marine Science Laboratory in Chapel Hill, North Carolina. Our approach is to generate internal solitary waves under stable continuous stratification in the modular wave tank of the Joint Fluids lab. The dimensions of our tank, $27$m long, up to $3$m high and $0.75$m wide (see figures \ref{CAD}(a,b)) allow us to reach larger scales than previously achieved in laboratory experiments, such as those reported in \citet{KoopB:1981,Grue:1999,Michallet:98,Kodaira:16,Carr:17}.

The modular tank was setup in two different ways: a long and shallow experiment where the wave was propagated down the full length of $27$m, and a deep but shorter experiment where the wave propagated over $9$m in the $2.25$m deep section. In the deep setup, the base layer was $2.15$m deep whereas the top layer was $10$cm.
In the case of long and shallow runs, the adjustable gate between the long and the deep sections remained fully open, for a total fluid depth of $77$ cm consisting of a top and bottom layer of $15$ cm and $62$ cm respectively (see figure \ref{CAD}a). This particular choice was made to match the experiments of \citet{Grue:1999} in order to proved a term of comparison for our experimental data with the shallow configuration. The laser and the imaging system, consisting of a double set of a Particle Image Velocimetry (PIV) camera and Laser Induced Fluorescence (LIF) cameras, all equipped with narrow band-pass optical filters -- respectively at $532$nm for the PIV and $583$nm for the LIF -- were mounted on a computer-controlled motorized cart. 
The full setup was either kept in a fixed position or towed along the tank to track the spatio-temporal evolution of the waves as they traveled down the long section (see figure \ref{CAD}b). 
In the case of the deep, short runs, the adjustable gate was closed to create a water tight seal between the deep and the long section. It will be shown later that these conditions are sufficient to generate nonlinear internal waves in the deep regime in a salt-stratified environment. 
The laser was mounted under the tank, and a single PIV/LIF setup was mounted on a fixed platform at the level of the top layer.
In both cases, an $80$cm tall gate formed an ``internal" dam that was either pulled up during the long runs or down during the deep runs in order to release the trapped fresh water.\\

\begin{figure}
\centering
\begin{center}
\begin{minipage}[b]{0.49\linewidth}
(a)\hspace{0mm}\scalebox{0.65}{\Large\input{Ini_prof_deep_2.tex}}
  \end{minipage}
  \begin{minipage}[b]{0.49\linewidth}
(b)\hspace{0mm}\scalebox{0.65}{\Large\input{Ini_prof_long_2.tex}}
  \end{minipage}
  \vspace{3mm}
  
(c)\includegraphics[scale=.33]{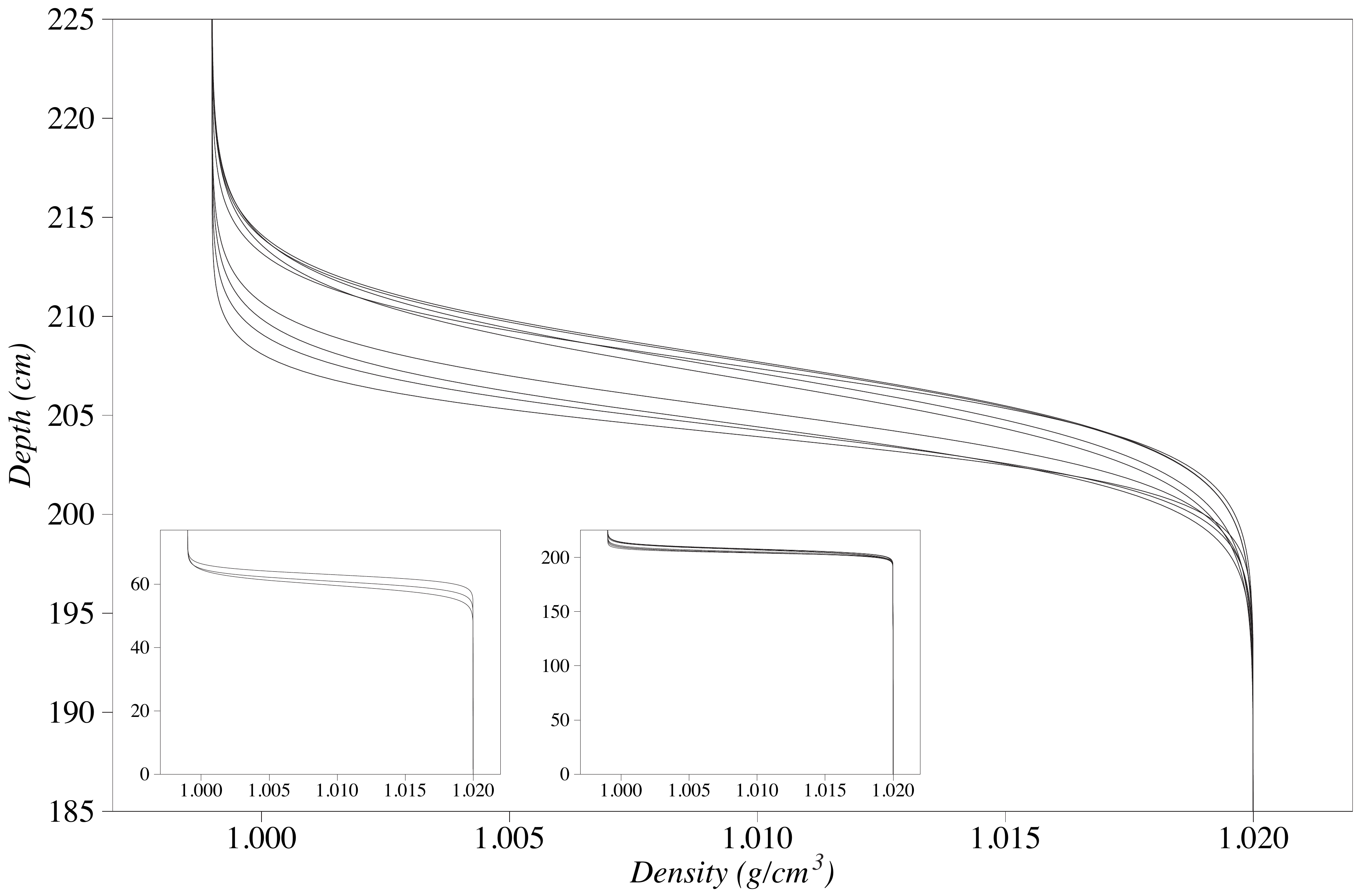} \\
\end{center}
\caption{Background density profiles extracted from the experiments  and used in both numerical simulations and DJL calculations. (a) Initial profiles normalized with respect to the reference profile $B(z)=(\tanh(4(x+1))-1)/2$ in the deep/short configuration and (b) in the long/shallow configuration.
(c) Details of the background stratification showing the variability of the height and thickness of the pycnocline between experiments for the deep regime. Insets: comparison between long/shallow configuration (left) and deep/short configuration (right). }
\label{numplots} 
\end{figure}

Stratifications were generated by combining salt and fresh water to a desired salinity, and then pumping it into the flume from the adjacent storage and processing unit. This allowed for profiles within the tank ranging from a nearly two-layer structure to continuous stratifications of general form. All tank sides, including the bottom,  consist of tempered glass panels, which allow for visualization from all directions. The flow was illuminated using a Litron$^\textsf{\textregistered}$ Nd:Yag diode pumped double cavity laser with $100$mJ per pulse at $532$nm wavelength used for both PIV and LIF (cf. figure \ref{CAD}(a,b)).
PIV combined with LIF data were collected with a custom built camera setup using up to four GigE Bobcat B3320 Imperx$^\textsf{\textregistered}$ $8$Mpx CCD cameras triggered simultaneously with the laser using a Stanford Research$^\textsf{\textregistered}$ DG535 pulse generator and acquired simultaneously on a computer using three Ethernet cards. The analysis of the experimental data was performed using the Matlab$^\textsf{\textregistered}$-based open-source PIV software DPIVSoft \citep{MeunierL:2003,PassaggiaLE:2012} to process the PIV and the LIF images. In addition to the stationary camera set-up, a moving camera mount followed, and moved slightly faster than, the wave in order to capture their evolutions over a longer distance.

Initial density measurements were acquired using an Orion TetraCon$^\textsf{\textregistered}$ 325 conductivity probe connected to a WTW Cond$^\textsf{\textregistered}$ 197I conductivity meter, which was then checked against the known bottom-layer salinity in parts-per-thousand, verifyied using a Anton$^\textsf{\textregistered}$ Paar DM35 densitometer whose calibration was verified to the fourth digit.
Initial density profiles were measured before each experiment using the Orion TetraCon$^\textsf{\textregistered}$ 325  conductivity probe mounted on a Velmex$^\textsf{\textregistered}$ linear stage. Measurements of conductivity and temperature were collected every centimeter and interpolated to calculate the corresponding density profile. These initial vertical profiles of density are shown in figure \ref{numplots}(a-c) for both the deep and the long section. In particular, with $\rho(z)$ the density at the vertical location $z$, and $\Delta \rho$ the total density variation from bottom to top (with reference density $\rho_0$ being the minimum density), we denote hereafter the renormalized density profile by $B(z)$, where \begin{equation}
B(z)=\frac{\rho(z)-\rho_0}{\Delta\rho};
\end{equation}
this is shown in figure \ref{numplots}(a) in the case of the deep configuration and \ref{numplots}(b) for the long/shallow  configuration.

The top layer was dyed with 0.066 $\mu$L per liters of water using BrightDyes$^\textsf{\textregistered}$ Rhodamine WT. Lavision$^\textsf{\textregistered}$ nearly neutrally buoyant polydisperse polyethylene PIV particles with diameters in the range [10:100] $\mu$m were mixed in a separate tank, sprayed over the free surface and left to slowly settle across the layers to seed the experiment. Note that the settling speed was several orders of magnitude smaller than the speed of the waves.
The laser was double-pulsed at $3$Hz while image pairs were recorded by mean of double exposure with a $20$ms difference between two successive frames.
The camera set-up was centered at $2.46$m from the end of the tank, or $15$m from the end of the long experiments. The purpose of this location was to capture a well-developed (i.e. sufficiently separated from the wake) solitary wave while minimizing end effects.

In our experimental facility, the salt water is recycled after each experiment. It is sent to a reverse osmosis unit which separates the salt component, and is subsequently filtered using a carbon and UV filters. The recycled water out of the reverse osmosis unit is then sent to $15$ storage tanks containing up to $23$m$^3$ of brine (six tanks) and fresh water (nine tanks). 




\section{Numerical methods}\label{sec:num_methods}
We consider the dynamics of an inviscid incompressible fluid
governed by the Euler equations, restricted to two-dimensions
which are written
\begin{subeqnarray}
    u_t + uu_x + wu_z &=& - \displaystyle \frac{p_x}{\rho}, \\
    w_t + uw_x + ww_z &=& - \displaystyle\frac{p_z}{\rho} - g, \\
    \rho_t + u\rho_x + w\rho_z &=& 0, \\
    u_x + w_z &=& 0.
\end{subeqnarray}
In the notation above, $(u,w)$ are the horizontal and vertical fluid velocities, respectively, $\rho$ is the (variable) fluid density, and $p$ is the pressure. Throughout the present manuscript, subscripts indicate partial derivatives. These equations are discretized and solved using the VarDen algorithm \citep{Almgren:1998} with initial conditions chosen to match those in the experiments. The numerical strategy of the VarDen code uses a second-order accurate projection method and a second-order predictor-corrector scheme for time integration. Both time-step and spatial mesh are adaptive with a base mesh discretization of $\Delta x \approx 0.44 \mbox{cm}$ and $\Delta x \approx 0.30 \mbox{cm}$ in the deep and long cases respectively.

The initial conditions for the dam-break are approximated using a hyperbolic tangent function profile given by
\begin{equation}
\rho(x,z) = 
\overline{\rho}\left[z-  \frac{H_{\rm{gate}}}{2}\big(1 + \tanh(W_{\rm{gate}} - x ) \big) \right],   
\end{equation}
where $\overline{\rho}(z)$ is the background density, $W_{\rm{gate}}$ the width of the gate (fixed at $56$ cm in both experiments and simulations) and $H_{\rm{gate}}$ is the depth of the fluid behind the gate beyond the rest state. The background density $\overline{\rho}$ is assumed to be a hyperbolic tangent function and it is best-fit to match the density profile measured before each experiment. Boundary conditions are taken to be free-slip, no-flux walls including the top of the domain; this rigid lid assumption could be viewed as the most significant departure of the simulations from the experiments. However, this is expected to cause  negligible effects  given the bounds on density differences, although our results do show a legacy of weak free-surface effects (cf. \S 4).

The dam-break method of generation and finite domain of the tank preclude true traveling waves from being observed experimentally. For point of comparison, however, we include the predictions of the DJL equation
\begin{equation}
	\nabla^2 \eta + \frac{N^2(z - \eta)}{c^2}\eta = 0,
\end{equation}
in which $\eta$ is the vertical displacement of a given isopycnal (constant $\rho$) line, $N^2(z - \eta) = -g\overline{\rho}'(z-\eta)/\overline{\rho}(z-\eta)$ is the Brunt-V\"ais\"al\"a frequency, and $c$ the speed of the traveling wave. The DJL equation is a reduction of the Euler equations to a steady waveframe $x \to x - ct$, and here we used the Boussinesq approximation version, whereby density variations are neglected for the inertial forces. The DJL equation has a natural variational formulation, so by using an iterative method that minimizes the energy  traveling wave solutions can be computed efficiently \citep{Stastna:02}. The present results are evaluated using the multi-grid strategy of \citet{Dunphy:11}, with the DJL solutions  obtained by continuation with increasing energy using fixed point iterations. 
The domain sizes for the DJL solver were $[L,H]=[9,2.15]$m for the deep case and $[L,H]=[9,0.77]$m for the long tank. The discretization consisted of equi-distributed points on a multi-grid strategy starting from $[n_x,n_z]=[256,256]$ for the coarse grid and up to $[n_x,n_z]=[2048,2048]$ in the case of the fine grid for the DJL solver. Convergence was also made possible starting with a thicker interface and was progressively decreased  through four levels of refinement down to the target thickness value.

The stratification at $x\rightarrow\pm L/2$ was kept constant and set by a hyperbolic type tangent of the form 
\begin{equation}
\overline{\rho}(z) = \rho_0 + \frac{\Delta\rho}{2}\left(1+ \tanh\left(\frac{z-z_0}{\delta} \right)\right),
\end{equation}
where $\rho_0=0.998$g/cc is the reference density of the fresh water, measured at room temperature at $T=23^o$C, $\Delta\rho$ is the maximum density difference measured between the top and the bottom of the tank, $\delta$ is the parameter that defines the thickness of the pycnocline, and $z_0$ is the height of the inflection point of the density profile. 

\begin{figure*}[t!]
\centering
(a)\hspace{0mm}\includegraphics[angle=0,width=120mm]{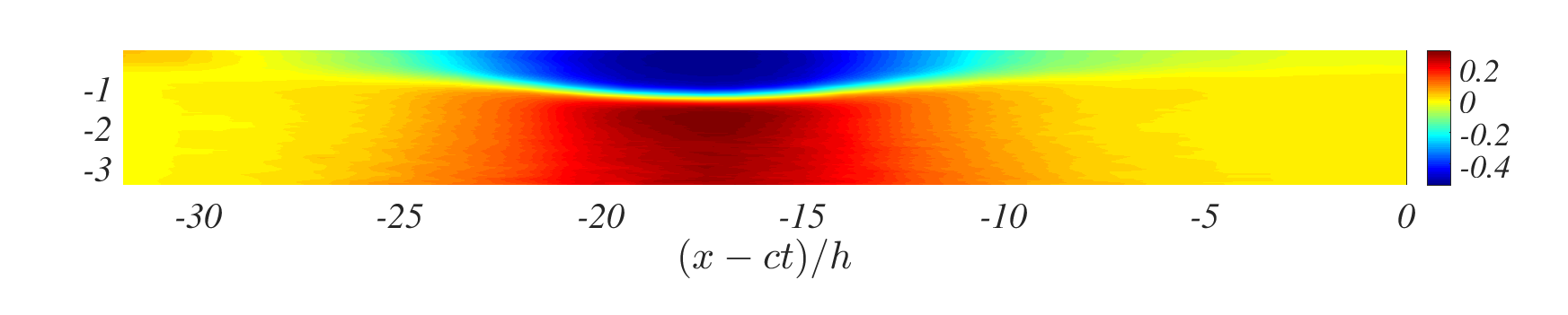}
\put(-337,28){\rotatebox{90}{\small $(H-z)/h$}}\\
(b)\hspace{0mm}\includegraphics[angle=0,width=120mm]{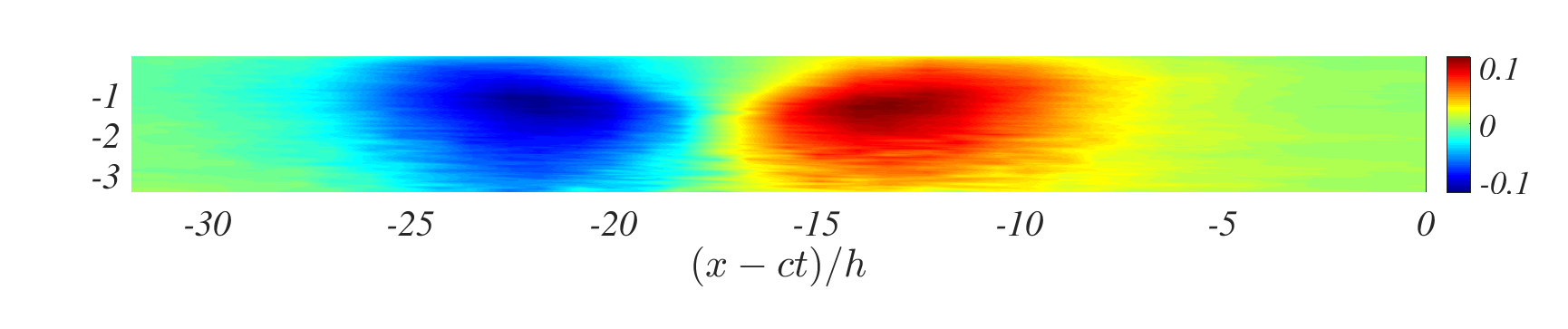}
\put(-337,28){\rotatebox{90}{\small $(H-z)/h$}}\\
(c)\hspace{0mm}\includegraphics[angle=0,width=120mm]{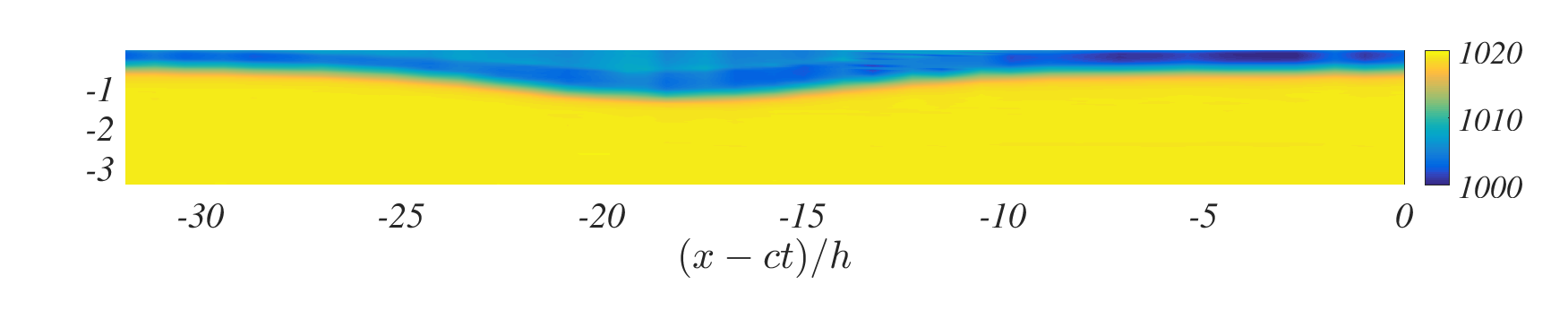}
\put(-337,28){\rotatebox{90}{\small $(H-z)/h$}}\\
(d)\hspace{0mm}\includegraphics[angle=0,width=120mm]{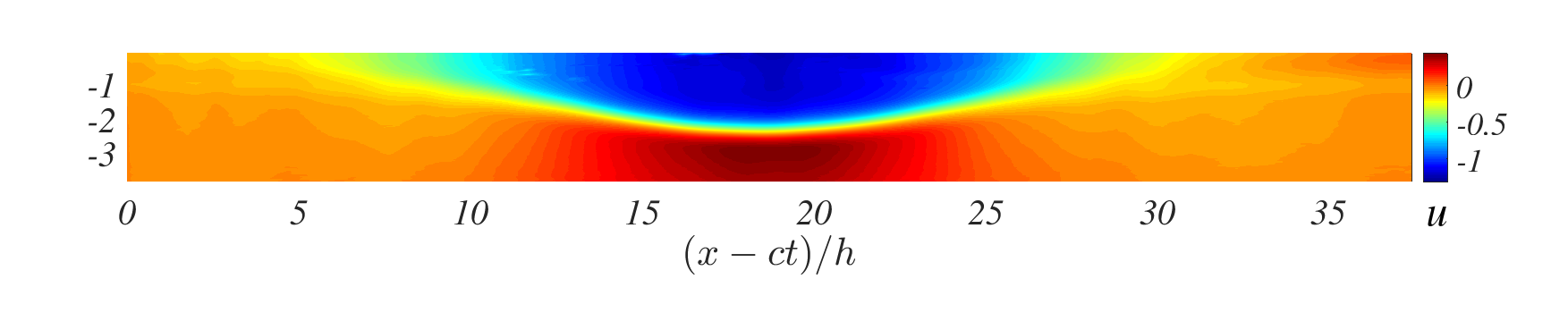}
\put(-337,28){\rotatebox{90}{\small $(H-z)/h$}}\\
(e)\hspace{0mm}\includegraphics[angle=0,width=120mm]{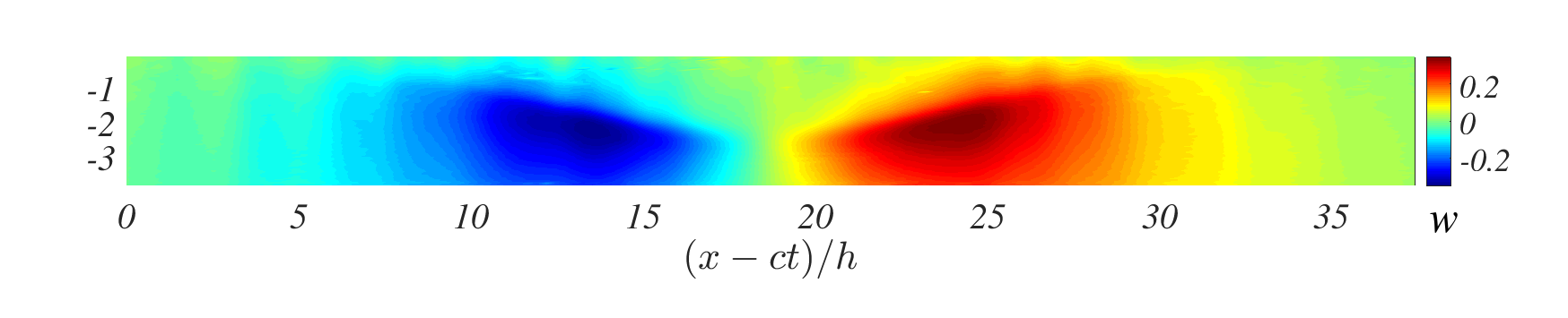}
\put(-337,28){\rotatebox{90}{\small $(H-z)/h$}}\\
(f)\hspace{0mm}\includegraphics[angle=0,width=120mm]{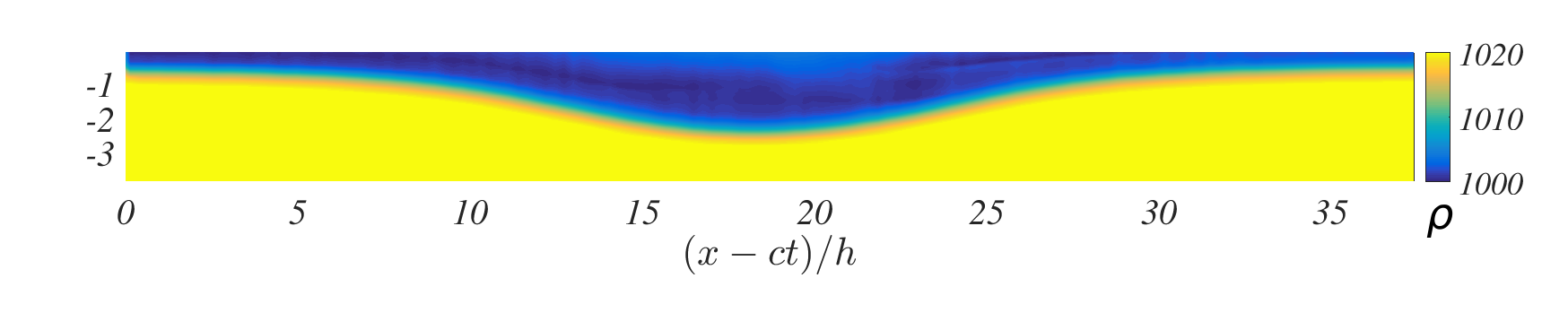}
\put(-337,28){\rotatebox{90}{\small $(H-z)/h$}}\\
\caption{Example of internal wave data in experiments for the long/shallow setup: (a-c) horizontal velocity field $u((x-ct)/h$ vs. $(H-z)/h)$, vertical velocity $w((x-ct)/h$ vs. $(H-z)/h)$ and density field $\rho((x-ct)/h$ vs. $(H-z)/h)$, respectively,  for a dam height $d/h=6\pm 10\%$, and a thin pycnocline $\delta/h\approx 0.55$, corresponding to $c/\sqrt{g'h}\approx1.43$, $a/h\approx0.718$ and $\lambda/h\approx5.04$. (d-f) Similar to (a-c) but for a dam height $d/h\approx2\pm 10\%$, and a thin pycnocline $\delta/h\approx 0.4$, corresponding to a wave with characteristics $c/\sqrt{g'h}\approx1.748$, $a/h\approx1.544$ and $\lambda/h\approx3.329$.}
\label{color_profiles} 
\end{figure*}

\section{Results}\label{sec:results}

Direct numerical simulations of the initial value problem for the dam-break-type initial solution are compared with experimental results and traveling-wave solutions computed from the DJL equations. In what follows, we first report a comparison for wave properties between the three methods, and show that onset of nonlinear internal solitary wave propagation is observed in deep water from fixed location measurements. Next, horizontal wave profiles from fixed measurements are analyzed for both deep and long configurations. Finally, in  \S\ref{track}, similar properties are examined but for data collected from a moving frame, whereby internal solitary waves are being tracked along the long section, in both the experiment and the simulation, to investigate their spatio-temporal evolution.

\subsection{Measurements from stationary locations}

\begin{figure}
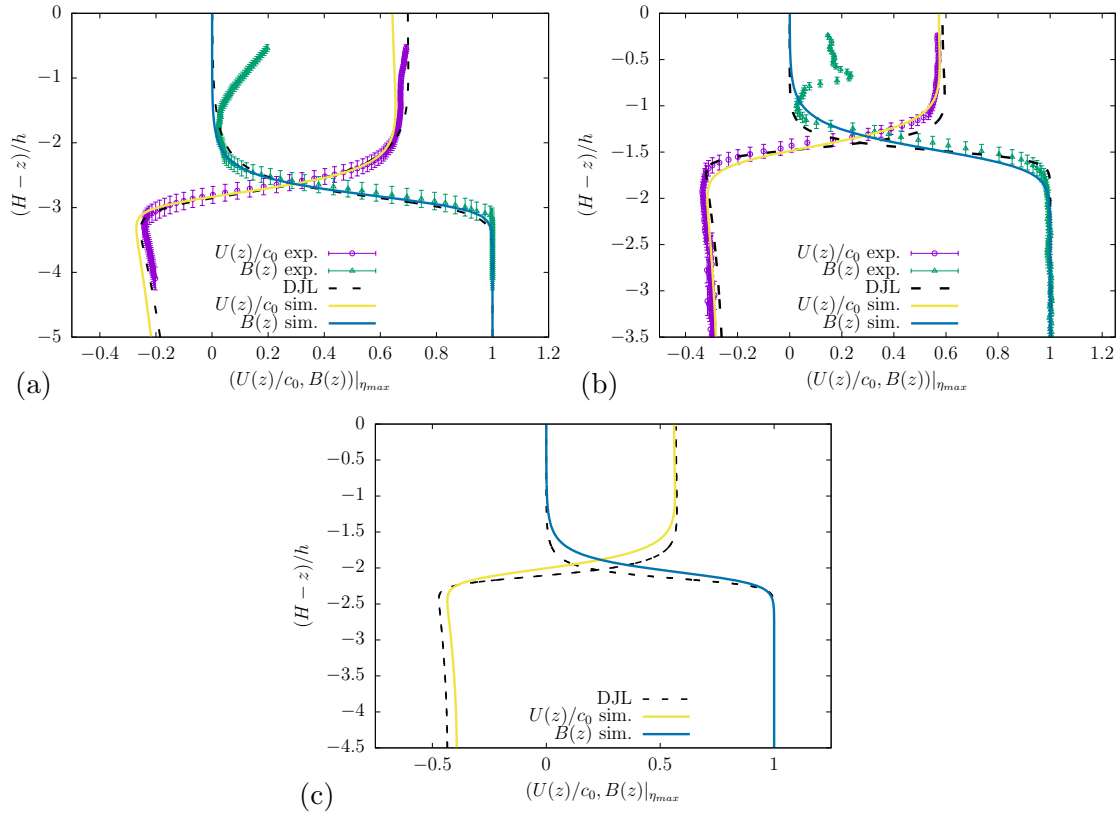

\centering
(a)\hspace{-6mm}\scalebox{0.6}{\large\input{U_B_Ri_z_0727_2.tex}}
(b)\hspace{-6mm}\scalebox{0.6}{\large\input{U_B_Ri_z_0915_2.tex}}\\
  (c)\hspace{-6mm}\scalebox{0.6}{\large\input{U_B_z_7m_2.tex}}
\caption{Example of velocity profiles $u(z)$ and density profiles $b(z)$ extracted from the PIV and LIF data at the center (maximum pycnocline displacement) of a wave, collected  along the wavetank at different distances $x$ from the generation gate; (b) long/shallow configuration, $x/h=85$ ; (c) long/shallow configuration,  $x/h=65$.
}
\label{profiles} 
\end{figure}

The results were extracted from the LIF and PIV data as illustrated in figure \ref{color_profiles}(a-c) for the deep setup and figure \ref{color_profiles}(d-f) for the long setup. The spatio-temporal evolution of the waves in  both the deep and long experiments are shown by means of density identified from the LIF, used as proxy and velocity computed using PIV. Figure \ref{color_profiles}(d-f) shows the largest amplitude wave generated in the long tank, and replicates the waves measured by \citet{Grue:1999}. The particular wave shown in figure \ref{color_profiles}(a) is compared with a much larger wave obtained in the deep tank in figure \ref{color_profiles}(c). 
The experiments in the long configuration were repeated
multiple times to check for the consistency of the generation mechanism,
using similar initial density profiles as shown in figure \ref{numplots}, and the results are displayed for both experimental data, DJL predictions and DNS simulations in figures \ref{plots}(a-d).
Instantaneous profiles of horizontal velocity and density (through the proxy by  LIF) are shown in figure \ref{profiles}(a) at $x/h=70$ from the dam gate for the deep case and figure \ref{profiles}(b) at $x/h=80$ for the long case.
These measurements are compared with the companion DNS of the initial value problem and the DJL results. The error measured between the simulation and the corresponding experiments at the location of maximum wave amplitude $\eta_{\rm{max}}$ is less than 10\% between the simulation and the DJL profile and 4\% between the experiment and the direct numerical simulation. Both the amplitude and the velocity are closely predicted for the deep and the long experiments. Here, the 
DJL solutions were computed by matching the amplitude with the experiment. The overall agreement is satisfactory, and the results, including measurements from other waves, show a slight improvement of the comparison with increasing amplitude.
Note that near the surface the density profile inferred from LIF data departs from the simulation, due to either light absorption by the dye (cf. figure \ref{profiles}(a)) or uneven concentration due to fading at this particular location in the tank (cf. figure \ref{profiles}(b)). The size of the facility, and the length of time for the set up of each experiment, made the dye intensity difficult to control. 

It is also interesting to notice that the vertical wave profile for the larger amplitude waves exhibits vertical oscillations, shown in figure \ref{color_profiles}(b), which appear to be synchronized with the oscillations of the free surface (not shown here). The wavelength of this oscillation is much shorter than the wavelength of the internal solitary wave, and seem to be dominant near the the front of the wave, a feature that agrees with  the oned recently reported by \citet{Kodaira:16} for large amplitude interfacial ISWs in immiscible fluids.

\subsection{Global ISWs characteristics}

Wave characteristics defined by amplitude $a$, wave speed $c$ and effective wavelength $\lambda/h$ are shown in figure \ref{numplots}(a-d), where the full symbols represent the results extracted from the two-dimensional Euler simulations, and the hollow symbols with error bars depict the experimental results.
All results are non-dimensionalized by the characteristic depth $h$ given by the mean-density isoline in the rest state, and the characteristic velocity $c_0^2 = g\left(\rho_{\rm{max}}/\rho_{\rm{min}}-1\right)h$. The wave speed $c$ was measured using two temporal mean-density displacement profiles located at fixed locations in the tank and separated by a distance $\Delta x \approx 25$cm apart from each other. Similarly to the experimental investigations of \citet{Luzzatto-FegizH14}, the velocity was computed by minimizing the time difference $\tau$ between the two temporal profiles, with the wave velocity calculated using $c\approx\Delta x/\tau$. Once determined, the speed of the wave was used to transform the temporal scale $t$ to the spatial scale $X$ through the transformation $X=x-ct$. The effective wavelength $\lambda/h$ is defined as
\begin{equation}
	\lambda = \frac{1}{a}\int_0^\infty \zeta(X) dX
\end{equation}
with $\zeta(X) = \zeta(x- ct)$ the displacement of the mean-density isoline, but computed in practice by integrating in time along the transformed spatial axis $X$ from the beginning of the experiment until the crest of the wave is observed in both the experiments and the numerical simulations.

\begin{figure*}[t!]
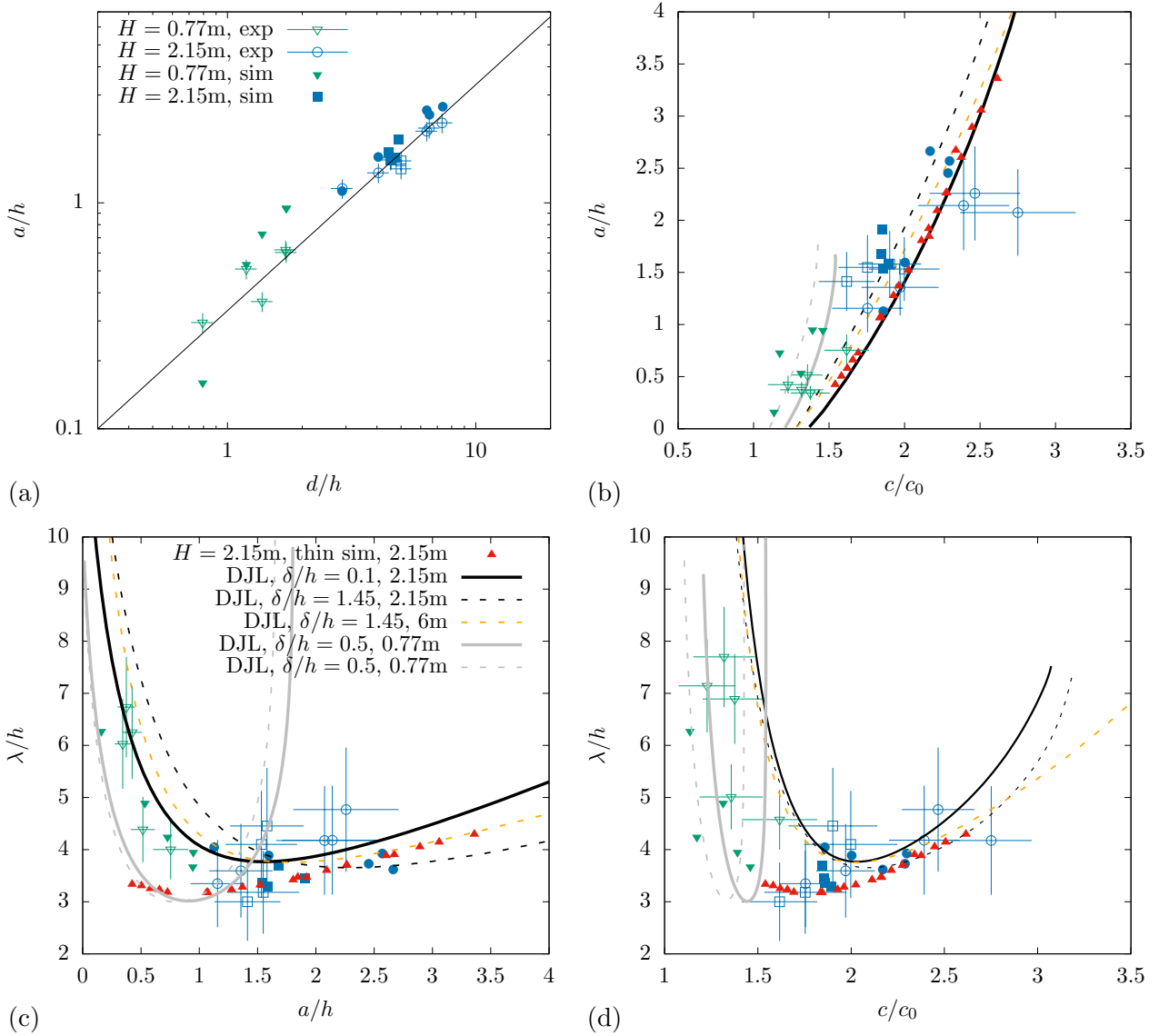

\centering
\begin{minipage}[b]{0.49\linewidth}
(a)\hspace{-6mm}\scalebox{0.85}{\input{a-h_d_2}}
  \end{minipage}
  \begin{minipage}[b]{0.49\linewidth}
(b)\hspace{-6mm}\scalebox{0.85}{\input{C_a-h_2.tex}}
  \end{minipage}
  \begin{minipage}[b]{0.49\linewidth}
(c)\hspace{-6mm}\scalebox{0.85}{\input{red-wave-length_a-h_2.tex}}
  \end{minipage}
  \begin{minipage}[b]{0.49\linewidth}
(d)\hspace{-6mm}\scalebox{0.85}{\input{red-wave-length_C_2.tex}}
  \end{minipage}
\caption{ISW characteristics, summary of all simulations and experiments: (a)
 ISW's amplitude $a/h$ as a function of the dam height $d/h$, (b) amplitude $a/h$ as function of wave speed $c/\sqrt{g'h}$, (c) reduced wavelength $\lambda/h$ as function of wave amplitude $a/h$, and (d) reduced wavelength $\lambda/h$ as function of wave speed $c/\sqrt{g'h}$. The full symbols show the numerical simulations, the hollow symbols with error bars refer to the experiments, and the solid and dashed curves are the theoretical predictions from the DJL equation at various locations and diffused interface thicknesses, as labeled in the legends.}
\label{plots} 
\end{figure*}

The amplitude, both experimentally and numerically,  of each ISW as a function of the dam height is shown in figure \ref{plots}(a), where the response of the ISW amplitude $a/h$ to increasing dam height $d/h$ remains essentially linear. 
This relationship indicates the consistency of the generation mechanism and the clear relation between dam height and resulting solitary wave amplitude. This confirms that the present experiments essentially lie within the deep regime, because no saturation of amplitude with dam height was observed, as it would be expected for the shallow regime when approaching a conjugate state.
The normalized effective wavelength $\lambda/h$ as a function of the wave speed is shown in figure \ref{plots}(b), which shows that  our measurements are well within the transition from linear to nonlinear internal solitary waves. The DJL solutions,  depicted  by means of solid and dashed black lines for the thicker and thinner pycnoclines in figures \ref{plots}(b-d), both describe a minimum effective wavelength $\lambda/h$ as a function of the wave speed $c/\sqrt{g'h}$.  The critical point found in the vicinity $[\lambda/h, c/\sqrt{g'h}]\approx[4,2]$ by the DJL computation is confirmed by the numerical simulations with a thin pycnocline (shown by means of red triangles (\textcolor{red}{$\blacktriangle$})). 
This critical point is also captured by the experiments and associated DNS, although the thicker pycnoclines suggest that this minimum shifts to slightly lower values of both $\lambda/h$ and $c/\sqrt{g'h}$. 

\begin{figure*}[t!]
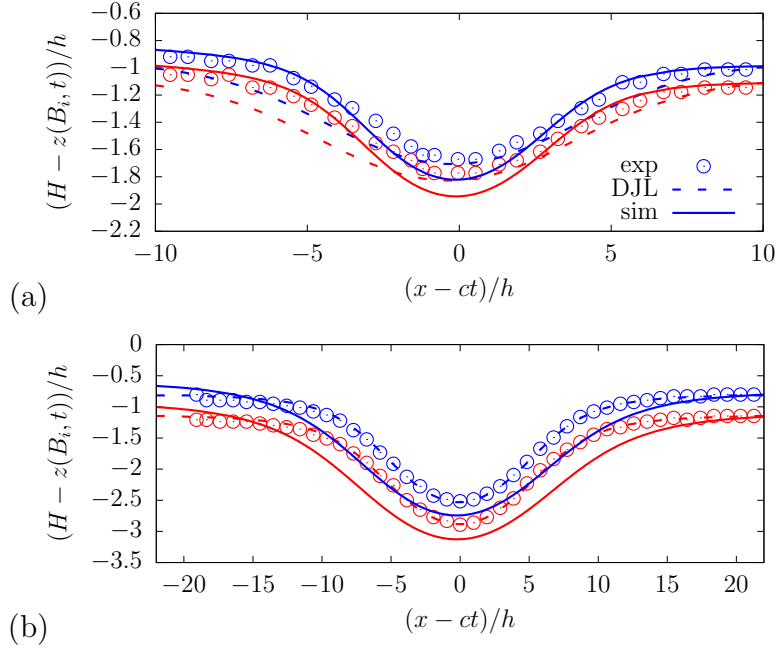

\centering\large
(a)\hspace{-1mm}\scalebox{0.8}{\input{seventy_thirty_long_2}}\\
(b)\hspace{-1mm}\scalebox{0.8}{\input{seventy_thirty_deep_2}}\\
\caption{Height of constant density $z(B_i,t)$, where $B_i$ is a given value with $i=[1,2]$, in  for a long experiment (a) and a deep experiment (b) in the time-space coordinate $(x-ct)/h$. The dashed lines are iso-density of the thirtieth (blue) and seventieth (red) percentile of the density, corresponding to $B_1=0.3$ and $B_2=0.7$ respectively.}
\label{wave_profiles} 
\end{figure*}

The normalized speed $c/\sqrt{g'h}$ is displayed as a function of the amplitude $a/h$ in figure \ref{plots}(b), where the DJL prediction is found to closely predict the speed simulated by the nearly two-layer stratification shown by the red triangles. The ISWs' speed are well within the bounds predicted by the DJL calculations except for the three larger amplitude experimental waves,  which appeared to be faster than both the DJL prediction and the respective numerical simulations. However a closer look at the PIV and LIF data shows that these waves were strongly asymmetric, with what appeared to be a recirculating core, from which large amplitude billows were observed developing downstream.

\begin{figure*}[t!]
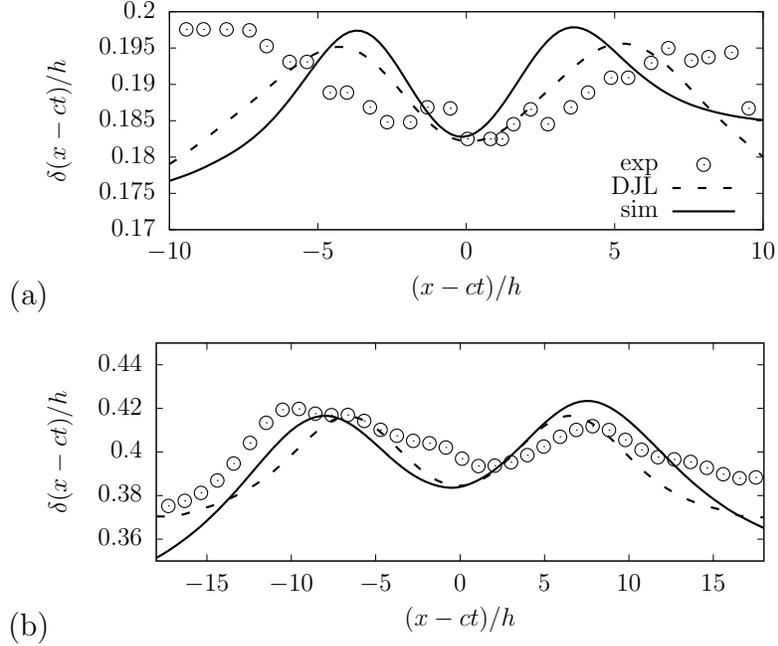

\centering\large
(a)\hspace{-1mm}\scalebox{0.8}{\input{thick_long_2}}\\
(b)\hspace{-1mm}\scalebox{0.8}{\input{thick_deep_2}}\\
\caption{Thickness $\delta(x-ct)$ of the diffused interface, measured by the difference between the thirtieth and seventieth percentile of the density reported in figures \ref{wave_profiles}(a,b) for: (a) the long configuration, and (b) the deep case in the time-space coordinate.}
\label{thickness} 
\end{figure*}

\begin{figure*}[t!]
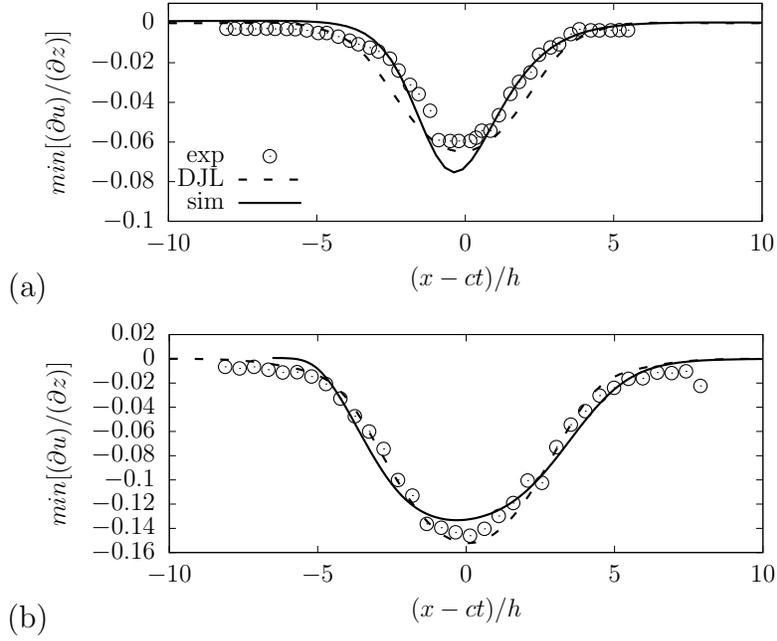

\centering\large
(a)\hspace{-1mm}\scalebox{0.8}{\input{shear_long_2}}\\
(b)\hspace{-1mm}\scalebox{0.8}{\input{Shear_deep_2}}\\
\caption{Minimum value of the shear ${\partial u(x-ct,z)}/{\partial z}$ computed along the waves reported in figures \ref{wave_profiles}(a,b) for the long (a) and deep (b) cases in the time-space coordinate.}
\label{shear} 
\end{figure*}

Figure \ref{plots}(c) depicts the effective wavelength $\lambda/h$ as a function of the ISWs' amplitude $a/h$, where a good collapse between theory, numerical simulations and experiments is further confirmed. The DJL results show the effect of pycnocline thickness, parametrized in nondimensional form by   
$\delta/h$, on the wave properties. As $\delta/h$ increases, the critical amplitude $a/h$ at which the minimum of the effective wavelength $\lambda(\delta/h)$ occurs also increases, while  $\lambda/h$ appears to remain essentially independent on the thickness in the range of  pycnoclines measured in our experiments, from $\delta/h=0.1$ for the thinnest to the thicker values of $\delta/h=1.45$.

Similar conclusions can be drawn from figure \ref{plots}(d) where the effective wavelength $\lambda/h$ is reported as a function of the non-dimensional wave speed $c/c_0$. The collapse between simulations and experiments is well within the DJL predictions where the thickness of the pycnocline does not seem to play an important role.

\subsection{Wave profiles}

The errors in the collapse between the DJL wave properties, the experiments and the numerical simulations suggest that the wave profiles provide more insights on wave details such as  symmetry and the effects of wakes. The DJL solutions of the Euler equations are  symmetric, traveling, steady states waves in the moving frame of reference $X \to x - ct$. In contrast, we are interested in comparing wave profiles for both experiments and numerical simulations, which are developing solitary waves that have not yet reached an asymptotic steady state (which of course could only be achieved adiabatically with respect to the slow decay due to the  presence of viscosity in real experiments). Profiles of solitary waves are shown in figure \ref{wave_profiles}(a) for the long configuration and figure \ref{wave_profiles}(b) for the deep experiments tracking isopycnals  from the experimental data, by matching the dye intensity with the corresponding value of the density. Values of the thirtieth and seventieth percentile of the density are reported in figure \ref{wave_profiles}(a,b), where the amplitude of the DJL solution was matched with the experimental results. The collapse between experiments and the DJL profile is particularly striking for the deep case. While the numerical simulation are in good agreement with DJL profiles, apart from a slight asymmetry with respect to the trough of the wave in figure \ref{wave_profiles}(b), the experimental and numerical results seem to degrade with respect to the DJL solution for the smaller amplitude wave shown in figure~\ref{wave_profiles}(a). Despite the small mismatch in amplitude, the asymmetry between the leading  and the trailing edge of the wave seems to be more important than for its large amplitude counterpart. The most likely scenario for this discrepancy seems to be related to the lack of separation from the wave's wake, wherein large amplitude, two-dimensional structures are seen to persist in the numerical simulations. This asymmetry is a consequence of the generation mechanism, which we show in the next section to be less apparent as the ISW travels farther along the wave tank. Of course, to increase the fidelity of the simulations with respect to experiments the former would have to be implemented in fully three-dimensional settings, substantially increasing the computation cost.

The difference between the two isopycnals difference $\Delta\delta/h$ above is shown in figure~\ref{thickness}(a,b) for both the deep and the long case, respectively, where we aim at confirming that the expansion and compression of the isopycnals induced by the wave can be observed in the experiment, and is captured by the DJL and direct numerical simulations. In the long/shallow case, figure \ref{thickness}(a) shows this comparison between the DJL solution, the simulation and the experiment. At the leading side of the wave, the isopycnals undergo an expansion, followed by a compression at the trough and a re-expansion along the lee side. The comparison between the DJL waves and the simulations is  in good qualitative agreement for both amplitude and locations of compression and expansion. The comparison with the experimental data is less precise, and it is barely captured for the experiments in the long configuration (see figure \ref{thickness}a) where the variations are only of the order of $\approx1.5$\% of $h$. The comparison and agreement in the deep case and a larger amplitude wave is more striking (see figure \ref{thickness}b). Variations in the pycnocline thickness are found to be of the order of $\approx 5\%$ of $h$ for the numerical simulation as well as for the DJL wave and the experiment.

Finally, we report the minimum value of the shear computed from the DJL solution, the experiment and the numerical simulation in figure \ref{shear}(a,b) for the short/deep and the long/shallow cases. The collapse between experiments, DJL solutions and numerical simulations is well predicted for both cases. This plot tests the quality of our experimental and numerical approaches. In the next section we extend the present analysis beyond the scales previously reported by \citet{Grue:1999} for the long/shallow regimes, and investigate the evolution of these ISWs by tracking their spatio-temporal evolution. To the best of our knowledge, this is the first time that such measurements with data collected in the co-moving frame of the wave has been attempted.
This helps determine the trend of our waves in eventually reaching a traveling (adiabatic) steady state form farther down the long section.

\subsection{Measurements from tracked ISWs}
\label{track}

\begin{figure*}[t!]
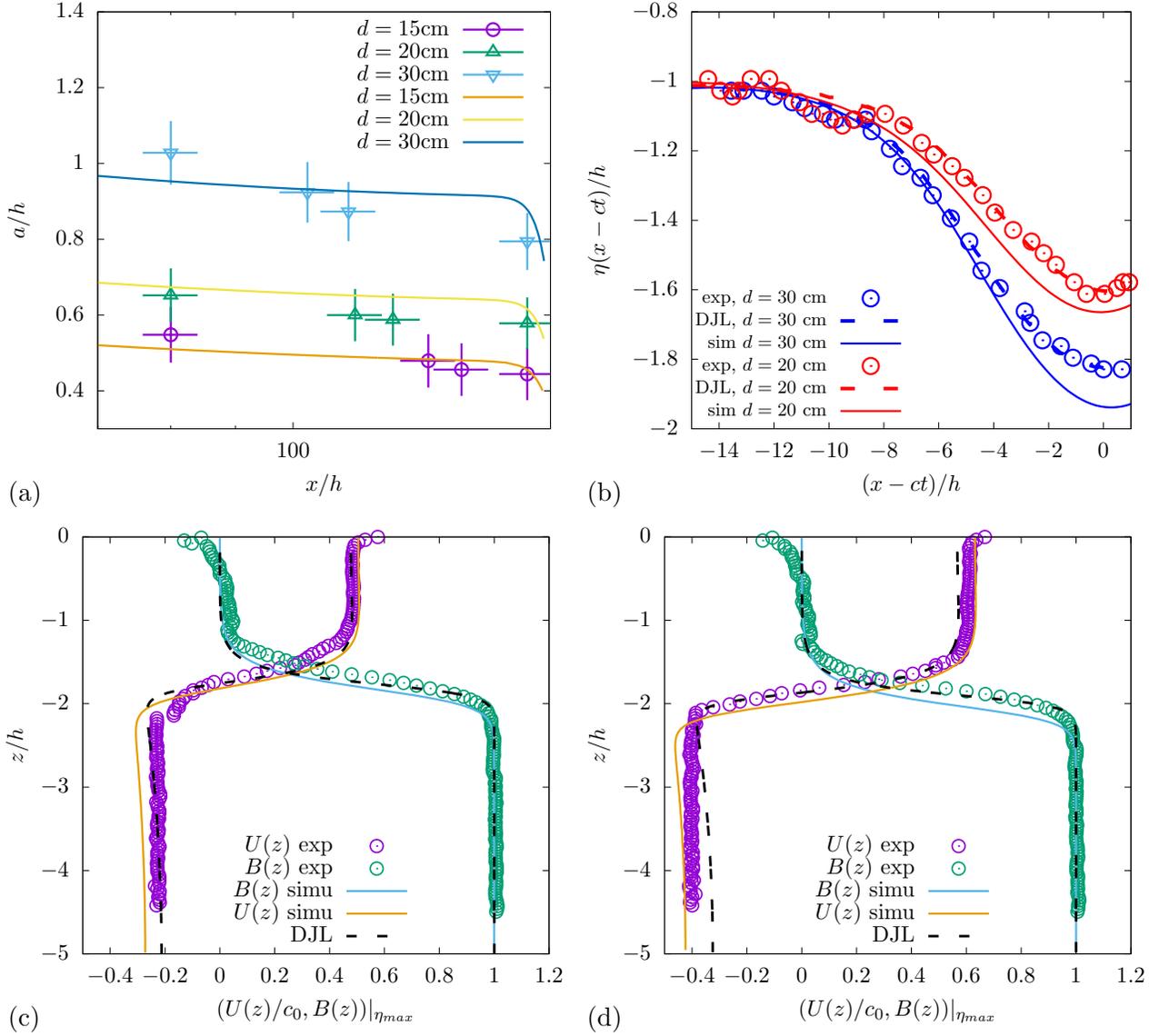

\centering
\begin{minipage}[b]{0.49\linewidth}
(a)\hspace{-6mm}\scalebox{0.85}{\input{a-x_d_2.tex}}
  \end{minipage}
  \begin{minipage}[b]{0.49\linewidth}
(b)\hspace{-6mm}\scalebox{0.85}{\input{fifty_end_long_2.tex}}
  \end{minipage}
  \begin{minipage}[b]{0.49\linewidth}
(c)\hspace{-6mm}\scalebox{0.85}{\input{U_B_z_end_small_2.tex}}
  \end{minipage}
  \begin{minipage}[b]{0.49\linewidth}
(d)\hspace{-6mm}\scalebox{0.85}{\input{U_B_z_end_large_2.tex}}
  \end{minipage}
\caption{Wave properties from moving frame: (a) Normalized ISW amplitude $a/h$ as a function of the non-dimensional cart position $x/h$ tracked in the long section for three representative dam height.
(b) Profiles of the fiftieth percentile of the density inferred from the LIF measured at $x/h\approx160$ from the dam for an initial dam heights. 
$(d-h)/h\approx1$ (red) and $(d-h)/h\approx2$ (blue). Profiles of horizontal velocity (purple-orange) measured by PIV (symbols) and normalized density $B=(\rho(z)-\rho_{min})/(\rho_{max}-\rho_{min})$ (blue-green) inferred from the LIF at the position of maximum displacement $\eta_{\rm{max}}$ compared with the DJL solution (lines) computed for the same amplitudes at $(d-h)/h\approx1.33$ (c) and $(d-h)/h\approx2$ (d).}
\label{end_tank_plots} 
\end{figure*}

The discrepancy for the measurements performed at $12$m from the gate in the long section led us to track several ISWs by following them with the moving  motorized cart shown on top of the long section in figure~\ref{CAD}. Once the ISWs were seen to pass the first location at $x=12$m, the cart was started with a velocity $10\%$ greater than the traveling wave speed for a KdV solitary wave \citep{Choi:1999}
\[
	c = c_0\left(1 - \frac{a}{2}\frac{\rho_1(H-h)^2 - \rho_2h^2}{\rho_1h(H-h)^2 + \rho_2h^2(H-h)} \right).
\]

Tracking ISWs remained however a difficult task due to the variability of wave speed immediately after the generation. 
We report only a limited number of measurements in figures \ref{end_tank_plots}(a-d) for 3 ISWs with initial dam heights $d=[15,20,30]$cm. 
The maximum amplitude, measured using both the LIF and the region of maximum shear in the PIV are reported in figure \ref{end_tank_plots}(a) as a function of the distance traveled $x/h$. The amplitude of these waves was found to decay slowly  as they travelled along the $27m$ of the long section of our modular tank;  this decay became less pronounced as the waves separated from their wakes. It is interesting to notice that both numerical simulations and experiments show a decay of the wave amplitude as the ISWs travel farther down the long section \citep{Grimshaw:03}. We remark that the simulation solves the Euler equation and does not include viscosity, the only source of dissipation being induced by the numerical scheme itself.
Also, we remark that the larger amplitude ISWs measured in the experiment seem to decay faster than the amount predicted by the simulation. This could be attributed to the vibrations induced by the cart carrying the laser, cameras, and power supplies, which can inject noise in the fluid flow, as discussed below.

Measurements of wave profiles inferred from the LIF are shown in figure~\ref{end_tank_plots}(b) for two representative amplitudes and compared with DJL solutions for the lee-side of the waves, where an overshoot was originally observed at $x/h=80$. The collapse between experiments and theory appears to improve significantly, except for the near-wake region. Here transient growth of finite amplitude noise, induced by the vibrations of the cart, may be responsible for the transition to Kelvin-Helmholtz-type instabilities which are seen as oscillations at both the trough, the lee and more strikingly in the wake of the wave \citep{CamassaV12,passaggia2018optimal}. Note that stationary measurements, which did not induce such vibrations, displayed no noticeable growing transient inside the wave region.

The good agreement in figure \ref{end_tank_plots}(b) suggests that the ISWs are becoming close to a traveling wave-type solution. To support this conclusion, profiles of density measured at the trough of the ISWs, inferred from the LIF for the density and horizontal velocity measured from PIV, are compared in figure \ref{end_tank_plots}(c,d) for two representative waves.
Both profiles appear to be in very good agreements between DJL and the simulations. Furthermore the overshoot observed at $x/h\approx80$ has now disappeared, and while horizontal velocity near the bottom show a slight mismatch of the order of $8\%$ -- probably due to finite tank effects -- the overall agreement is excellent. This suggests that our ISWs have to travel a distance of $x/h \approx 120$ from their generation site in order to more closely approach a traveling wave-type state.

\begin{figure*}[t!]
\begin{center}
(a)\hspace{0mm}\scalebox{0.85}{\input{red-wave-length_a-h_KandB_2}}\\
(b)\hspace{-1mm}\scalebox{0.9}{\input{C_a-h_all_2.tex}}
\end{center}
\caption{Effective wave length $\lambda/h$ as a function of wave amplitude $a/h$ for the present results in salt=stratified water ompared with previous experimental results from \citet{KoopB:1981} (black symbols) for the case of immiscible fluids (water and freon),  the KdV prediction (thin dashed line (- - -)), two-layer model \citet{Grue:97} (thin dashed-dotted line $-\cdot-\cdot-$), fully nonlinear model ({\bf-----}) \citet{Choi:1999,Choi:1996} in the long/shallow case, the field data of \citet{Liu04} (large hollow triangles), \citet{Ramp04} (large hollow circles) and \citet{Huang16} (large hollow pentagon). Thick lines are the DJL predictions shown in figure \ref{plots}(c). (b) same as figure \ref{plots}(c) but compared with field data from the above references.}
\label{waves} 
\end{figure*}

The possible application of our results in the context of open ocean measurements and theoretical models is further discussed next, together with our conclusions on the present study.

\section{Discussion}\label{sec:discussion}

The present study reports a clear experimental evidence of nonlinear ISWs in the deep regime and miscible fluids. Results are systematically compared with DJL solutions, direct numerical simulations of the dam-break initial value problem, and the associated experiments. Measurements are reported for the fully nonlinear regime in a deep configuration, past the critical point, which is confirmed by the DJL theory where the nondimensional effective wavelength increases with respect to both the wave amplitude and wave speed. These results are supported with local measurements of LIF and PIV, which are further compared with the DJL waves and the direct numerical simulations with an overall satisfactory agreement.
In addition, we explore the effect of the interface thickness on the dynamics, and show that thick pycnoclines modify both wave amplitude and  wave speed. Larger thicknesses increase the wave amplitude while decreasing the speed and the wavelength, an effect that does not seem to be experimentally studied in the same depth of other properties of internal waves.

Vertical and horizontal wave profiles are also compared between numerical simulations, experiments and DJL solutions. For these properties, such velocity and density measurements, the agreement is excellent.
Horizontal profiles of isopycnal displacement demonstrate the overall good agreement between the DJL solutions, numerical simulations and experiments. We show in the long/shallow configuration that waves measured too close to their region of generation tend to be asymmetric, as they have not yet fully separated from their wakes. This effect is however not observed farther down the long section and shows that relatively large experimental facilities are necessary in order to observe the approach to true traveling waves in the mathematical sense of models such as KdV,  DJL, and Camassa-Choi  \citep{korteweg1895xli,dubreil1933determination,Choi:1999}. Such asymmetries have also been reported in oceanic observations and are known to play an important role in the redistribution of nutrients and chlorophyll, which are essential drivers of primary production in the euphotic zone \citep{dong2015asymmetry}.

The present results are gathered in figure \ref{waves}(a) together with the results from \citet{KoopB:1981} where interfacial ISWs were generated in immiscible fluids (water and liquid freon) and the linear/nonlinear theories for the deep and shallow cases. In addition, we report large amplitude ISWs wave characteristics from field measurements in the deep regime, in the case of a train of internal solitary waves propagating westward off the Oregon Shelf with very thin stratification \citep{Stanton:98} (\textcolor{green}{\Large{$\diamond$}}). We also compare with ISWs 
propagating westward from Luzon straight in the South China Sea from \citet{Liu04} 
(\textcolor{magenta}{\Large{$\circ$}}) 
and \citet{Ramp04} 
(\textcolor{red}{\Large{$\triangle$}}), 
and an extreme wave developing eastward in the Northern China Sea \citep{Huang16} 
(\textcolor{cyan}{\Large{$\pentagon$}}).
Effective wavelengths $\lambda/h$ for both DJL, experimental, and field measurements are reported in figure~\ref{waves}(a) as a function of wave amplitude $a/h$. As displayed by this figure, our measurements fill the gap between the two-layer experiments of \citet{KoopB:1981} and \citet{Michallet:98} and field measurements of large amplitude ISWs in deep waters.
At this point, a remark is in order:  while experiments producing nonlinear ISWs could be performed in tabletop set-ups, our experiments bridge the gap between tabletop experiments (subject to small-scale effects such as viscosity, surface tension, and Marangoni convection at the surface \citep{Luzzatto-FegizH14} and field observations. Instead, we observed large-scale effects such as transient growth and free-surface motions.
The consistency of results across length-scales suggests that laboratory ISW characteristics faithfully represent ISWs measured from field observations.
Wave speeds from field measurements are also compared with the DJL solutions, experimental data, and numerical results in figure \ref{waves}(b), which shows an overall good agreement. These data support the idea that large amplitude ISWs in the nonlinear regime from field observations in the ocean \citep{Stanton:98,Huang16} or in lakes \citep{Preusse2012seasonal,Preusse:12} are only accurately modelled by DJL type solutions \citep{Dunphy:11} and nonlinear theories such as the Choi-Camassa two-layer model \citep{Choi:1999}.


PIV results and images from experiments indicate the presence of small surface waves traveling with the ISW for the larger amplitude ISWs. The resonance between capillary and gravity waves has been recently explored experimentally and theoretically for interfacial ISWs in immiscible fluids by \citet{Kodaira:16}. Our numerical code is based on a rigid lid approximation, and thus cannot accommodate free surface motions. The present experimental results could however provide an interesting starting point to study surface-internal wave interactions in the case of miscible fluids, which deserves more investigation and represents an important application for the dynamics and prediction of large amplitude internal solitary waves \citep{Alford:15}.

\subsection*{Acknowledgements}
RC, RMM, and CT acknowledge the support by the National Science Foundation under grants RTG DMS-0943851, CMG ARC-1025523, DMS-1009750, DMS-1517879, and DURIP N00014-12-1-0749. PYP acknowledges the support by the National Science Foundation under grant NSF OCE-1155558 and NSF OCE-1736989. RC, RMM, and CT thank David Adalsteinsson for helpful comments on the post-processing of the numerical results using DataTank.

\bibliographystyle{jfm}

\bibliography{bib}

\begin{thebibliography}{58}
\expandafter\ifx\csname natexlab\endcsname\relax\def\natexlab#1{#1}\fi

\bibitem[Alford {\em et~al.\/}(2015)Alford, Peacock, MacKinnon, Nash, Buijsman,
  Centuroni, Chao, Chang, Farmer, Fringer {\em et~al.\/}]{Alford:15}
{\sc Alford, M.~H., Peacock, T., MacKinnon, J.~A., Nash, J.~D., Buijsman,
  M.~C., Centuroni, Luca~R, Chao, S.-Y., Chang, M.-H., Farmer, D.~M., Fringer,
  O.~B. {\em et~al.\/}} 2015 The formation and fate of internal waves in the
  south china sea. {\em Nature\/} {\bf 521}~(7550), 65--69.

\bibitem[Almgren {\em et~al.\/}(1998)Almgren, Bell, Colella, Howell \&
  Welcome]{Almgren:1998}
{\sc Almgren, A.S., Bell, J.~B., Colella, P., Howell, L.~H. \& Welcome, M.L.}
  1998 A conservative adaptive projection method for the variable density
  incompressible navier--stokes equations. {\em J. Comput. Phys.\/} {\bf
  142}~(1), 1--46.

\bibitem[Apel {\em et~al.\/}(1985)Apel, Holbrook, Liu \& Tsai]{Apel:85}
{\sc Apel, J.~R., Holbrook, J.~R., Liu, A.~K. \& Tsai, J.~J.} 1985 The sulu sea
  internal soliton experiment. {\em J. Phys. Oceanogr.\/} {\bf 15}~(12),
  1625--1651.

\bibitem[Benjamin(1986)]{Benjamin:1986}
{\sc Benjamin, T.~B.} 1986 On the boussinesq model for two-dimensional wave
  motions in heterogeneous fluids. {\em J. Fluid Mech.\/} {\bf 165}, 445--474.

\bibitem[Bourgault {\em et~al.\/}(2016)Bourgault, Galbraith \&
  Chavanne]{bourgault:16}
{\sc Bourgault, D., Galbraith, P.~S. \& Chavanne, C.} 2016 Generation of
  internal solitary waves by frontally forced intrusions in geophysical flows.
  {\em Nature Commun.\/} {\bf 7}, 13606.

\bibitem[Brandt {\em et~al.\/}(2002)Brandt, Rubino \& Fischer]{brandt:02}
{\sc Brandt, Peter, Rubino, Angelo \& Fischer, J{\"u}rgen} 2002 Large-amplitude
  internal solitary waves in the north equatorial countercurrent. {\em J. Phys.
  Oceanogr.\/} {\bf 32}~(5), 1567--1573.

\bibitem[Camassa \& Viotti(2012)]{CamassaV12}
{\sc Camassa, R. \& Viotti, C.} 2012 On the response of large-amplitude
  internal waves to upstream disturbances. {\em J. of Fluid Mech.\/} {\bf 702},
  59--88.

\bibitem[Carr {\em et~al.\/}(2017)Carr, Franklin, King, Davies, Grue \&
  Dritschel]{Carr:17}
{\sc Carr, M., Franklin, J., King, S.~E., Davies, P.~A., Grue, J. \& Dritschel,
  D.~G.} 2017 The characteristics of billows generated by internal solitary
  waves. {\em J. Fluid Mech.\/} {\bf 812}, 541--577.

\bibitem[Carr {\em et~al.\/}(2008)Carr, Fructus, Grue, Jensen \&
  Davies]{Carr:08}
{\sc Carr, M., Fructus, D., Grue, J., Jensen, A. \& Davies, P.~A.} 2008
  Convectively induced shear instability in large amplitude internal solitary
  waves. {\em Phys. Fluids\/} {\bf 20}~(12), 126601.

\bibitem[Carr {\em et~al.\/}(2011)Carr, King \& Dritschel]{Carr:11}
{\sc Carr, M., King, S.~E. \& Dritschel, D.~G.} 2011 Numerical simulation of
  shear-induced instabilities in internal solitary waves. {\em J. Fluid
  Mech.\/} {\bf 683}, 263--288.

\bibitem[Choi \& Camassa(1996)]{Choi:1996}
{\sc Choi, W. \& Camassa, R.} 1996 Long internal waves of finite amplitude.
  {\em Phys. Rev. Lett.\/} {\bf 77}~(9), 1759.

\bibitem[Choi \& Camassa(1999)]{Choi:1999}
{\sc Choi, W. \& Camassa, R.} 1999 Fully nonlinear internal waves in a
  two-fluid system. {\em J. Fluid Mech.\/} {\bf 396}, 1--36.

\bibitem[Churnside \& Ostrovsky(2005)]{churnside:05}
{\sc Churnside, JH \& Ostrovsky, LA} 2005 Lidar observation of a strongly
  nonlinear internal wave train in the gulf of alaska. {\em Int. J. Remote
  Sens.\/} {\bf 26}~(1), 167--177.

\bibitem[Davis \& Acrivos(1967)]{DavisA:67}
{\sc Davis, R.~E. \& Acrivos, A.} 1967 Solitary internal waves in deep water.
  {\em J. Fluid Mech.\/} {\bf 29}~(3), 593--607.

\bibitem[Dong {\em et~al.\/}(2015)Dong, Zhao, Chen, Meng, Shi \&
  Tian]{dong2015asymmetry}
{\sc Dong, J., Zhao, W., Chen, H., Meng, Z., Shi, X. \& Tian, J.} 2015
  Asymmetry of internal waves and its effects on the ecological environment
  observed in the northern south china sea. {\em Deep Sea Res.\/} {\bf 98},
  94--101.

\bibitem[Dubreil-Jacotin(1933)]{dubreil1933determination}
{\sc Dubreil-Jacotin, M.-L.} 1933 Sur la d{\'e}termination rigoureuse des ondes
  permanentes p{\'e}riodiques d'ampleur finie. {\em C. R. Acad. Sci.\/} {\bf
  197}, 818.

\bibitem[Dubreil-Jacotin(1934)]{dubreil1934determination}
{\sc Dubreil-Jacotin, M.-L.} 1934 Part {I}: Sur la d{\'e}termination rigoureuse
  des ondes permanentes p{\'e}riodiques d'ampleur finie. {P}art {II}: Suite de
  composition dans la th{\'e}orie des groupes finis et abstraits et dans la
  th{\'e}orie des id{\'e}aux de p{\^o}lynomes. PhD thesis, {\'E}cole Normale
  Sup{\'e}rieur.

\bibitem[Duda {\em et~al.\/}(2004)Duda, Lynch, Irish, Beardsley, Ramp, Chiu,
  Tang \& Yang]{Duda:04}
{\sc Duda, T.~F., Lynch, J.~F., Irish, J.~D., Beardsley, R.~C., Ramp, S.~R.,
  Chiu, C.-S., Tang, Tswen~Y. \& Yang, Y.-J.} 2004 Internal tide and nonlinear
  internal wave behavior at the continental slope in the northern south china
  sea. {\em IEEE J. Ocean. Eng.\/} {\bf 29}~(4), 1105--1130.

\bibitem[Dunphy {\em et~al.\/}(2011)Dunphy, Subich \& Stastna]{Dunphy:11}
{\sc Dunphy, M., Subich, C. \& Stastna, M.} 2011 Spectral methods for internal
  waves: indistinguishable density profiles and double-humped solitary waves.
  {\em Nonlin. Proc. Geophys.\/} {\bf 18}~(3), 351--358.

\bibitem[Fructus {\em et~al.\/}(2009)Fructus, Carr, Grue, Jensen \&
  Davies]{Fructus:09}
{\sc Fructus, D., Carr, M., Grue, J., Jensen, A. \& Davies, P.~A.} 2009
  Shear-induced breaking of large internal solitary waves. {\em J. Fluid
  Mech.\/} {\bf 620}, 1--29.

\bibitem[Grimshaw(1981)]{Grimshaw:81}
{\sc Grimshaw, R.} 1981 Slowly varying solitary waves in deep fluids. In {\em
  Proc. Royal Soc. A\/}, , vol. 376, pp. 319--332. The Royal Society.

\bibitem[Grimshaw \& Helfrich(2012)]{grimshaw2012effect}
{\sc Grimshaw, R. \& Helfrich, K.} 2012 The effect of rotation on internal
  solitary waves. {\em IMA J. App. Math.\/} {\bf 77}~(3), 326--339.

\bibitem[Grimshaw {\em et~al.\/}(2003)Grimshaw, Pelinovsky \&
  Talipova]{Grimshaw:03}
{\sc Grimshaw, R., Pelinovsky, E. \& Talipova, T.} 2003 Damping of
  large-amplitude solitary waves. {\em Wave Motion\/} {\bf 37}~(4), 351--364.

\bibitem[Grue {\em et~al.\/}(1997)Grue, Friis, Palm \& Rus{\aa}s]{Grue:97}
{\sc Grue, J., Friis, H.~A., Palm, E. \& Rus{\aa}s, P.~O.} 1997 A method for
  computing unsteady fully nonlinear interfacial waves. {\em J. Fluid Mech.\/}
  {\bf 351}, 223--252.

\bibitem[Grue {\em et~al.\/}(1999)Grue, Jensen, Rus{\aa}s \& Sveen]{Grue:1999}
{\sc Grue, J., Jensen, A., Rus{\aa}s, P.-O. \& Sveen, J.~K.} 1999 Properties of
  large-amplitude internal waves. {\em J. Fluid Mech.\/} {\bf 380}, 257--278.

\bibitem[van Haren(2013)]{vanHaren:13}
{\sc van Haren, H.} 2013 Bottom-pressure observations of deep-sea internal
  hydrostatic and non-hydrostatic motions. {\em J. Fluid Mech.\/} {\bf 714},
  591--611.

\bibitem[Helfrich \& Melville(2006)]{HelfrichM:06}
{\sc Helfrich, K.~R. \& Melville, W.~K.} 2006 Long nonlinear internal waves.
  {\em Annu. Rev. Fluid Mech.\/} {\bf 38}, 395--425.

\bibitem[Hosegood \& van Haren(2006)]{Hosegood:06}
{\sc Hosegood, P. \& van Haren, H.} 2006 Sub-inertial modulation of
  semi-diurnal currents over the continental slope in the faeroe-shetland
  channel. {\em Deep Sea Res.\/} {\bf 53}~(4), 627--655.

\bibitem[Huang {\em et~al.\/}(2016)Huang, Chen, Zhao, Zhang, Zhou, Yang \&
  Tian]{Huang16}
{\sc Huang, X., Chen, Z., Zhao, W., Zhang, Z., Zhou, C., Yang, Q. \& Tian, J.}
  2016 An extreme internal solitary wave event observed in the northern south
  china sea. {\em Sci. Rep.\/} {\bf 6}, 30041.

\bibitem[Johnston {\em et~al.\/}(2015)Johnston, Rudnick \& Kelly]{Johnston:15}
{\sc Johnston, T. M.~S., Rudnick, D.~L. \& Kelly, S.~M.} 2015 Standing internal
  tides in the tasman sea observed by gliders. {\em J. Phys. Oceanogr.\/} {\bf
  45}~(11), 2715--2737.

\bibitem[Kalisch \& Bona(2000)]{KalischB:00}
{\sc Kalisch, Henrik \& Bona, Jerry~L} 2000 Models for internal waves in deep
  water. {\em Discrete and Continuous Dynamical Systems\/} {\bf 6}~(1), 1--20.

\bibitem[Kodaira {\em et~al.\/}(2016)Kodaira, Waseda, Miyata \&
  Choi]{Kodaira:16}
{\sc Kodaira, T., Waseda, T., Miyata, M. \& Choi, W.} 2016 Internal solitary
  waves in a two-fluid system with a free surface. {\em J. Fluid Mech.\/} {\bf
  804}, 201--223.

\bibitem[Koop \& Butler(1981)]{KoopB:1981}
{\sc Koop, C.~G. \& Butler, Ge.} 1981 An investigation of internal solitary
  waves in a two-fluid system. {\em J. Fluid Mech.\/} {\bf 112}, 225--251.

\bibitem[Korteweg \& De~Vries(1895)]{korteweg1895xli}
{\sc Korteweg, D.~J. \& De~Vries, G.} 1895 On the change of form of long waves
  advancing in a rectangular canal, and on a new type of long stationary waves.
  {\em London, Edinburgh \& Dublin Phil Mag. \& J. Sci.\/} {\bf 39}~(240),
  422--443.

\bibitem[Kubota {\em et~al.\/}(1978)Kubota, Ko \& Dobbs]{Kubota:78}
{\sc Kubota, T., Ko, D. R.~S. \& Dobbs, L.~D.} 1978 Weakly-nonlinear, long
  internal gravity waves in stratified fluids of finite depth. {\em AIAA J.
  Hydron.\/} {\bf 12}~(4), 157--165.

\bibitem[Kunze {\em et~al.\/}(2012)Kunze, MacKay, McPhee-Shaw, Morrice, Girton
  \& Terker]{Kunze:12}
{\sc Kunze, Eric, MacKay, Chris, McPhee-Shaw, Erika~E, Morrice, Katie, Girton,
  James~B \& Terker, Samantha~R} 2012 Turbulent mixing and exchange with
  interior waters on sloping boundaries. {\em J. Phys. Oceanogr.\/} {\bf
  42}~(6), 910--927.

\bibitem[Lamb(2014)]{Lamb:14}
{\sc Lamb, K.~G.} 2014 Internal wave breaking and dissipation mechanisms on the
  continental slope/shelf. {\em Ann. Rev. Fluid Mech.\/} {\bf 46}, 231--254.

\bibitem[Lien {\em et~al.\/}(2014)Lien, Henyey, Ma \& Yang]{Lien:14}
{\sc Lien, R.-C., Henyey, F., Ma, B. \& Yang, Y.~J.} 2014 Large-amplitude
  internal solitary waves observed in the northern south china sea: properties
  and energetics. {\em J. Phys. Oceanogr.\/} {\bf 44}~(4), 1095--1115.

\bibitem[Liu {\em et~al.\/}(1985)Liu, Holbrook \& Apel]{Liu:85}
{\sc Liu, A.~K., Holbrook, J.~R. \& Apel, J.~R.} 1985 Nonlinear internal wave
  evolution in the sulu sea. {\em J. Phys. Oceanogr.\/} {\bf 15}~(12),
  1613--1624.

\bibitem[Liu {\em et~al.\/}(2004)Liu, Ramp, Z. \& Tang]{Liu04}
{\sc Liu, A.~K., Ramp, S.~R., Z., Y. \& Tang, T.~Y.} 2004 A case study of
  internal solitary wave propagation during asiaex 2001. {\em IEEE J. Ocean.
  Eng.\/} {\bf 29}~(4), 1144--1156.

\bibitem[Long(1953)]{long1953some}
{\sc Long, R.~R.} 1953 Some aspects of the flow of stratified fluids: I. {A}
  theoretical investigation. {\em Tellus\/} {\bf 5}~(1), 42--58.

\bibitem[Luzzatto-Fegiz \& Helfrich(2014)]{Luzzatto-FegizH14}
{\sc Luzzatto-Fegiz, P. \& Helfrich, K.~R.} 2014 Laboratory experiments and
  simulations for solitary internal waves with trapped cores. {\em J. of Fluid
  Mech.\/} {\bf 757}, 354--380.

\bibitem[Meunier \& Leweke(2003)]{MeunierL:2003}
{\sc Meunier, P. \& Leweke, T.} 2003 Analysis and treatment of errors due to
  high velocity gradients in particle image velocimetry. {\em Exp. Fluids\/}
  {\bf 35}~(5), 408--421.

\bibitem[Michallet \& Barth\'elemy(1998)]{Michallet:98}
{\sc Michallet, H. \& Barth\'elemy, E.} 1998 Experimental study of interfacial
  solitary waves. {\em J. Fluid Mech.\/} {\bf 366}, 159--177.

\bibitem[Moum {\em et~al.\/}(2003)Moum, Farmer, Smyth, Armi \& Vagle]{Moum:03}
{\sc Moum, J.~N., Farmer, D.~M., Smyth, W.~D., Armi, L. \& Vagle, S.} 2003
  Structure and generation of turbulence at interfaces strained by internal
  solitary waves propagating shoreward over the continental shelf. {\em J.
  Phys. Oceanogr.\/} {\bf 33}~(10), 2093--2112.

\bibitem[Passaggia {\em et~al.\/}(2018)Passaggia, Helfrich \&
  White]{passaggia2018optimal}
{\sc Passaggia, P.-Y., Helfrich, K.~R. \& White, B.~L.} 2018 Optimal transient
  growth in thin-interface internal solitary waves. {\em J. Fluid Mech.\/} {\bf
  840}, 342--378.

\bibitem[Passaggia {\em et~al.\/}(2012)Passaggia, Leweke \&
  Ehrenstein]{PassaggiaLE:2012}
{\sc Passaggia, P.-Y., Leweke, T. \& Ehrenstein, U.} 2012 Transverse
  instability and low-frequency flapping in incompressible separated boundary
  layer flows: an experimental study. {\em J. Fluid Mech.\/} {\bf 703},
  363--373.

\bibitem[Pinkel(1979)]{Pinkel:79}
{\sc Pinkel, R.} 1979 Observations of strongly nonlinear internal motion in the
  open sea using a range-gated doppler sonar. {\em J. Phys. Oceanogr.\/} {\bf
  9}~(4), 675--686.

\bibitem[Preusse {\em et~al.\/}(2012{\natexlab{{\em a\/}}})Preusse,
  Freist{\"u}hler \& Peeters]{Preusse2012seasonal}
{\sc Preusse, Martina, Freist{\"u}hler, Heinrich \& Peeters, Frank}
  2012{\natexlab{{\em a\/}}} Seasonal variation of solitary wave properties in
  lake constance. {\em J. Geophys. Res. Oceans\/} {\bf 117}~(C4).

\bibitem[Preusse {\em et~al.\/}(2012{\natexlab{{\em b\/}}})Preusse, Stastna,
  Freist{\"u}hler \& Peeters]{Preusse:12}
{\sc Preusse, M., Stastna, M., Freist{\"u}hler, H. \& Peeters, F.}
  2012{\natexlab{{\em b\/}}} Intrinsic breaking of internal solitary waves in a
  deep lake. {\em PloS One\/} {\bf 7}~(7), e41674.

\bibitem[Ramp {\em et~al.\/}(2004)Ramp, Tang, Duda, Lynch, Liu, Chiu, Bahr, Kim
  \& Yang]{Ramp04}
{\sc Ramp, S.~R., Tang, T.~Y., Duda, T.~F., Lynch, J.~F., Liu, A.~K., Chiu,
  C.-S., Bahr, F.~L., Kim, H.-R. \& Yang, Y.-J.} 2004 Internal solitons in the
  northeastern south china sea. part i: Sources and deep water propagation.
  {\em IEEE J. Ocean. Eng.\/} {\bf 29}~(4), 1157--1181.

\bibitem[Rudnick {\em et~al.\/}(2003)Rudnick, Boyd, Brainard, Carter, Egbert,
  Gregg, Holloway, Klymak, Kunze, Lee {\em et~al.\/}]{Rudnick:03}
{\sc Rudnick, D.~L., Boyd, T.~J., Brainard, R.~E., Carter, G.~S., Egbert,
  G.~D., Gregg, M.~C., Holloway, P.~E., Klymak, J.~M., Kunze, E., Lee, C.~M.
  {\em et~al.\/}} 2003 From tides to mixing along the hawaiian ridge. {\em
  Science\/} {\bf 301}~(5631), 355--357.

\bibitem[Simmons \& Alford(2012)]{simmons2012simulating}
{\sc Simmons, H.~L. \& Alford, M.~H.} 2012 Simulating the long-range swell of
  internal waves generated by ocean storms. {\em Oceanogr.\/} {\bf 25}~(2),
  30--41.

\bibitem[Stanton \& Ostrovsky(1998)]{Stanton:98}
{\sc Stanton, T.~P. \& Ostrovsky, L.~A.} 1998 Observations of highly nonlinear
  internal solitons over the {C}ontinental {S}helf. {\em Geophys. Res. Lett.\/}
  {\bf 25}, 2695--2698.

\bibitem[Stastna \& Lamb(2002)]{Stastna:02}
{\sc Stastna, M. \& Lamb, K.~G.} 2002 Large fully nonlinear internal solitary
  waves: The effect of background current. {\em Phys. Fluids\/} {\bf 14}~(9),
  2987--2999.

\bibitem[Vlasenko {\em et~al.\/}(2000)Vlasenko, Brandt \&
  Rubino]{vlasenko2000structure}
{\sc Vlasenko, Vasiliy, Brandt, Peter \& Rubino, Angelo} 2000 Structure of
  large-amplitude internal solitary waves. {\em J. Phys. Oceanogr.\/} {\bf
  30}~(9), 2172--2185.

\bibitem[Xie {\em et~al.\/}(2015)Xie, He, Chen, Xu \& Cai]{xie2015simulations}
{\sc Xie, Jieshuo, He, Yinghui, Chen, Zhiwu, Xu, Jiexin \& Cai, Shuqun} 2015
  Simulations of internal solitary wave interactions with mesoscale eddies in
  the northeastern south china sea. {\em J. Phys. Oceanogr.\/} {\bf 45}~(12),
  2959--2978.

\bibitem[Zhao {\em et~al.\/}(2016)Zhao, Ertekin, Duan \& Webster]{Ertekin16}
{\sc Zhao, B.~B., Ertekin, R.~C., Duan, W.~Y. \& Webster, W.~C.} 2016 New
  internal-wave model in a two-layer fluid. {\em J. Waterway Port. Coastal
  Ocean Eng.\/} {\bf 142}, 04015022.

\end{thebibliography}

\end{document}